\documentclass{kapproc} 

\def\lesssim{\mathrel{\hbox{\rlap{\hbox{\lower4pt\hbox{$\sim$}}}\hbox{$<$}}}}
\def\gtrsim{\mathrel{\hbox{\rlap{\hbox{\lower4pt\hbox{$\sim$}}}\hbox{$>$}}}}
\newcommand{\energyinput}{$\dot{E}/A_{\rm SF}$}

\usepackage{procps} 

\usepackage[dvips]{graphicx}

\chapsectrunningheads

\upperandlowercase

 \let\footnote\savefootnote

\kluwerbib 

\begin{document}

\articletitle
{Radio halos in star forming galaxies}

\chaptitlerunninghead{Radio halos in star forming galaxies} 

\author{Ute Lisenfeld,\altaffilmark{1} Michael Dahlem,\altaffilmark{2}
\& Matthias Ehle\altaffilmark{3}}

\affil{\altaffilmark{1}Instituto de Astrof\'\i sica de Andaluc\'\i a (CSIC),
Granada, Spain \ 
\altaffilmark{2} ATNF, Narrabi, Australia \
\altaffilmark{3} XMM-Newton Science Operations Centre, ESA, Madrid, Spain
}

\begin{abstract}
We study the relation between radio halos, the energy input
by supernovae in the disk and the
galaxy mass. We find that both the energy input
by supernovae as well as 
the galaxy mass are important parameters for understanding the
 formation of  radio halos.
Galaxies with a high energy input
by supernovae per star forming area and a low galaxy mass  
generally possess radio halos whereas galaxies with the
opposite characteristics do not.
Furthermore, there is a tentative correlation
between the observed scale height
and the expected height in a simple gravitational approximation.

\end{abstract}

\section{Introduction}
 
There is accumulating observational evidence for the existence
of gaseous halos around disk galaxies (see \cite{det92} 1992 and 
\cite{dah97} 1997
for reviews), consisting of warm and hot ionized gas, dust,  
magnetic fields and cosmic rays (CRs), the latter two generating
the radio continuum
(synchrotron) emission. 
Theoretical models have been developed
to explain these halos, such as the galactic
fountain model (\cite{sha76} 1976), galactic chimneys 
(\cite{nor89} 1989), superwinds (\cite{hec90} 1990) and
superbubble outbreaks (\cite{mac99} 1999). All models are based on the
assumption that the energy source driving the 
formation of halos
are supernova (SN) explosions.

Observationally, it is still a matter of controversy how many and exactly
which galaxies have such halos. In order to answer this question,
we have been  observing radio halos
(\cite{dah95} 1995, \cite{dah01} 2001) in an ongoing project.
In the present paper we summarize some 
results and use existing data  to try to understand the formation of 
radio halos.

\section{Observations of radio halos}

\cite{dah01} (2001) observed a sample  of galaxies with
active star formation (SF) 
(selected with respect to the  IRAS flux ratio
between 60 and 100 $\mu$m,  $f_{60}/f_{100} \gtrsim 0.4$) 
with the VLA and the Australia Telescope Compact Array.
They fitted the radio emission perpendicular
to the disk with two exponential functions
convolved with the beam, modeling in this way a thin disk and a halo.
Radio halos were found in 6 out of 11 galaxies (55\%) with exponential scale
heights between 1.4 and 3.1 kpc.  
The high detection rate in these actively star forming
galaxies showed that SF is indeed a key factor for the
formation of radio halos. 
Possible reasons for the non-detection of halos in 5 galaxies could be:
(i) There is no radio halo. (ii) The resolution of the data is too low.
We are currently obtaining higher resolution data for some
objects. (iii) The width of the thin disk is much smaller than
the beam and therefore not resolved. 
This is supported by the fact that
for galaxies with no radio halo the apparent scale height of the
thin disk was much larger (between 1 and 2 kpc) than in galaxies
with a radio halo (less than 1 kpc).

In order to base our present  study
on a larger number of galaxies, we use, in addition to the 11 galaxies
from \cite{dah01} (2001), data for
6 galaxies described in \cite{dah95} (1995), 
as well as for
3 galaxies from \cite{irw99} (1999) who 
observed 16 edge-on galaxies searching for radio halos.
Their analysis of the radio emission is different
from ours and involves a ranking of the visibility of 
extended emission based on different criteria.
We include only those galaxies for which
their VLA D-array data does not  show evidence
of emission beyond their modelled thin disk (their Fig. 1),
a  robust criterion that is similar
to the one used in the rest of the sample.

\section{Understanding radio halos}

From theoretical considerations we expect that the formation of
radio halos depends (at least) on the following 
factors.

1) According to all models (see above) 
a fundamental parameter is the energy input  by SN explosions 
into the interstellar  medium.
\cite{dah95} (1995) showed that radio halos do not form above the
entire disk, but only out to radial distances where SF
takes place. Therefore, 
the relevant parameter is expected to be
the energy input by SNe per SF disk area, \energyinput.

2) The energy input from SNe allows material to be lifted above the
disk against the gravitational potential determined by the
mass of the galaxy. Therefore, the mass of the galaxy is expected to
play an important 
role.

3)  CR electrons have
a limited life-time due to inverse Compton and synchrotron
energy losses. The observed steepening of the synchrotron spectrum
with increasing distance from the disk (e.g. \cite{hum91} 1991)
is due to these energy losses
and shows that they are indeed important.
These energy losses limit the distance to which CR electrons
can travel and therefore the size of radio halos.
In the present work, we do
not take into account CR energy losses, because it would 
require a detailed knowledge of the distribution
of the energy density of the radiation field, $U_{\rm rad}$, (determining
the inverse Compton losses) and the magnetic field structure (causing the
synchrotron losses and emission and determining the CR propagation).
Qualitatively, we expect CR energy losses
to decrease the range of observed scale heights because galaxies with a 
high surface brightness (implying a high SF rate and thus a high
\energyinput) also possess a high $U_{\rm rad}$
causing important inverse Compton losses.

\begin{figure}[ht]
\centerline{\includegraphics[width=6.cm]{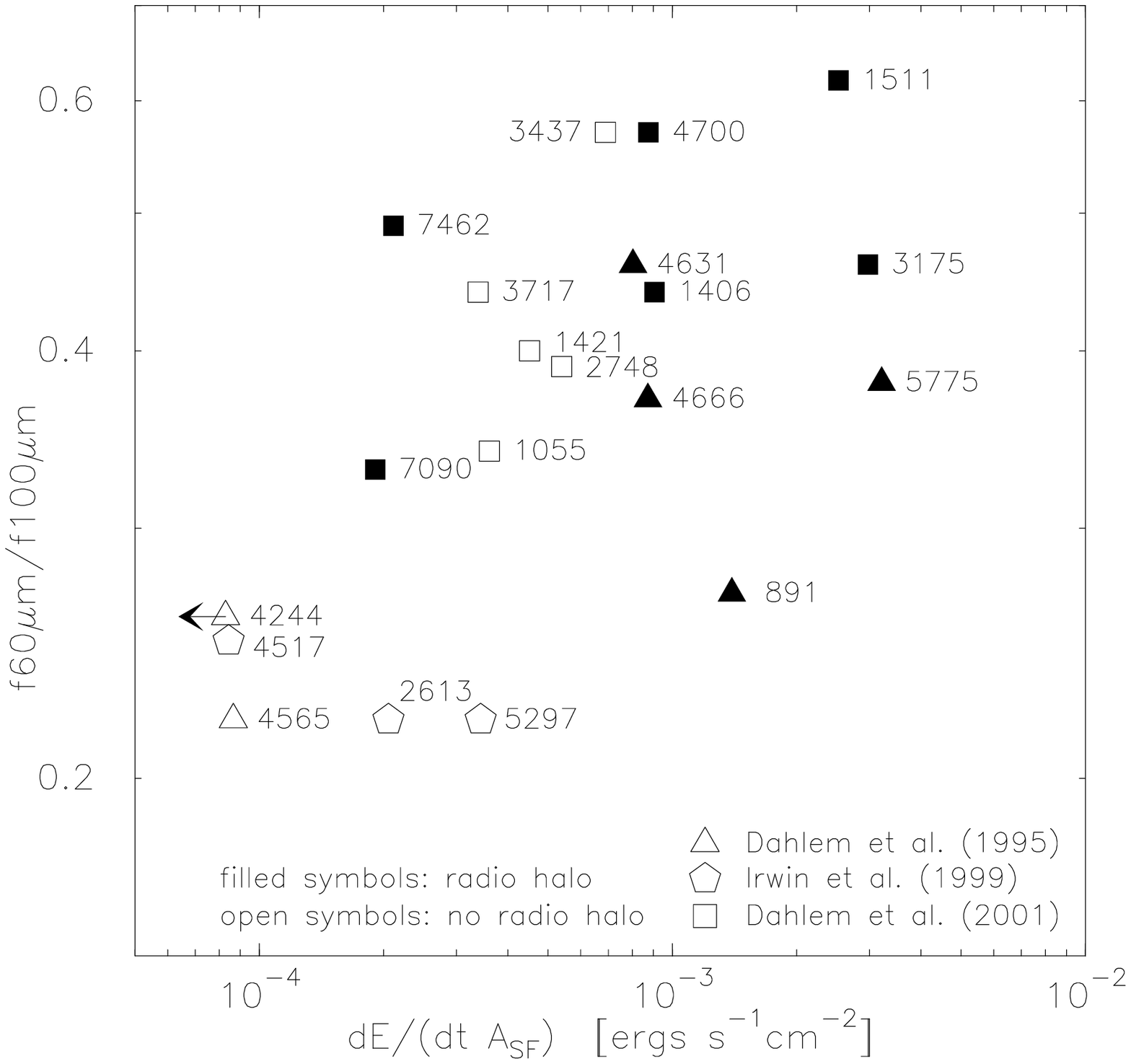}
\includegraphics[width=6.cm]{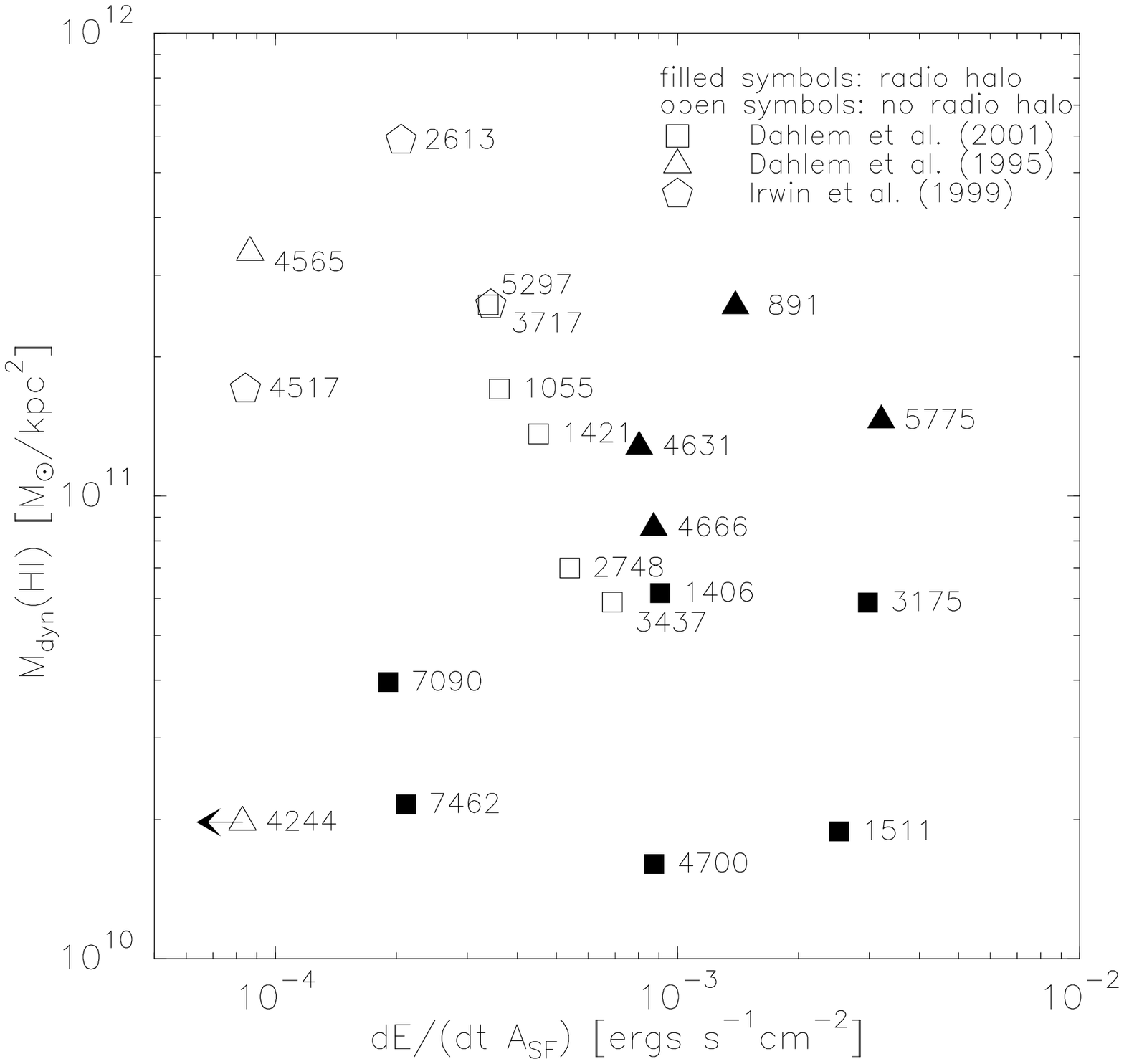}}
\caption{The energy input by SNe per SF area as a function of
the IRAS flux ratio between  60 and 100
$\mu$m, $f_{60}/f_{100}$ {\bf (left)},
and the dynamical mass of a galaxy, $M_{\rm dyn}$(HI), calculated from 
the HI linewidth {\bf (right)}. 
As a measure for the energy input by SNe we use the 
radio continuum emission as described
in \cite{dah95} (1995). 
}
\end{figure}

Fig. 1 (left) shows the dust temperature (measured as $f_{60}/f_{100}$),
which is an empirical measure for the SF activity,
as a function of the energy input by SNe per SF area.
There is a rough correlation between both quantities and a 
trend for radio halos to be found in galaxies with a high dust temperature,
respectively a high \energyinput.
A similar conclusion was drawn from the smaller sample used in
\cite{dah01} (2001) and by \cite{ros03} (2003) for a sample
of 74 edge-on galaxies observed in H$\alpha$.

Fig. 1 (right) shows \energyinput \
versus the dynamical mass of a galaxy.
There is a clear division between galaxies with and without
a halo: Galaxies with a low mass
and a high energy input (lower right side) have radio halos whereas 
galaxies with a high mass and a low energy input 
(upper left side) do not. 
This shows that the
galaxy mass plays an important role in the formation of radio halos.

\begin{figure}[ht]
\centerline{\includegraphics[width=6.cm]{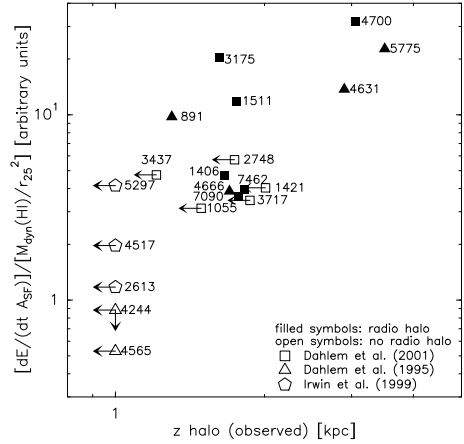}}
{\caption{The observed exponential scale height of the radio halos
versus the expected height
in the approximation described in the text.
For the galaxies without a detected radio halo from \cite{irw99} (1999)
and \cite{dah95} (1995) we adopted an arbitrary but
realistic upper limit of 1 kpc. For the galaxies from \cite{dah01}
(2001) without a detected radio halo we adopted 
the measured scale height of the thin disk as an upper limit.}}
\end{figure}

Fig. 2  shows the observed exponential scale heights  of the radio emission
versus the expected height in the simple approximation  
that the energy  input per mass
is proportional to \energyinput, that
energy losses of CR electrons are neglected and that the gravitational
potential is described by that of an infinite disk. With these
assumptions,
the expected height of the radio halo is proportional to 
\energyinput \ per mass surface density.
%
A trend is visible in the sense that
large radio halos are present in galaxies
with high values of (\energyinput) per mass surface density
whereas galaxies with no 
radio halos have a low values.
%
This indicates that -- although more detailed modelling taking into
account CR propagation is necessary for a  full understanding 
-- the energy input by SNe  and the
galaxy mass are important parameters to understand 
the properties of radio halos.
 
\begin{acknowledgments}
UL is partially supported
by the  Spanish MCyT  Grant  AYA 2002-03338 and
Junta de Andalucia (Spain).
\end{acknowledgments}

\end{document}